\documentclass[twocolumn,showpacs,showkeys,prl]{revtex4}
\usepackage{graphicx}
\usepackage{dcolumn}
\usepackage{bm}
\usepackage{subfigure}
\usepackage{hyperref}

\begin{document}


\title{Landau and Ott scaling for the kinetic energy density and
the low $T_c$ conventional superconductors, $Li_{2}Pd_{3}B$ and Nb.}

\author{Mauro M. Doria}
\author{S. Salem-Sugui Jr.}%
\affiliation{Instituto de Fisica, Universidade Federal do Rio de Janeiro, 21945-970 Rio de Janeiro, Brazil}
\author{P. Badica}
\affiliation{High Field Lab. for Superconducting Materials, Institute for Materials Research, Tohoku University,
2-1-1 Katahira, Aoba-ku, Sendai, 980-8577 Japan} \altaffiliation[Also at ]{National Institute of Materials
Physics, POB MG-7, 077125 Bucharest, Romania}
\author{K. Togano}
\affiliation{National Institute of Materials Science, Tsukuba, 1-2-1 Sengen, 305-0047 Japan.}

\date{\today}

\begin{abstract}
The scaling approach  recently proposed by Landau and Ott for isothermal magnetization curves is extended to the
average kinetic energy density of the condensate. Two low $T_c$ superconductors, Nb and $Li_{2}Pd_{3}B$ are
studied and their isothermal reversible magnetization shown to display Landau and Ott scaling. Good agreement is
obtained for the upper critical field $H_{c2}(T)$, determined from the Abrikosov approximation for the
reversible region (standard linear extrapolation of the magnetization curve), and from the maximum of the
kinetic energy curves. For the full range of data, which includes the irreversible region, the isothermal
$d.M.B/H^2$ curves for $Li_2Pd_3B$ show an impressive collapse into a single curve over the entire range of
field measurements. The Nb isothermal $d.M.B/H^2$ curves exhibit the interesting feature of a constant and
temperature independent minimum value.
\end{abstract}
\maketitle
Many methods have been used to determine the upper critical field $H_{c2}(T)$, some of them based on scaling
laws which are expected to hold for the reversible magnetization \cite{ullah,clem,LO}. Recently Landau and Ott
applied successfully a new scaling method to many high-$T_c$ reversible magnetization curves \cite{LO,LO2,LO3}.
Several high-$T_c$ materials\cite{LO2} data sets, obtained using different experimental procedures, were shown
by Landau and Ott to obey their predicted scaling. According to their scaling the $HvsT$ phase diagram border is
retrieved just from one single isothermal curve $MvsH$ taken at some temperature T. M and H refer to the
magnetization and the applied field, respectively. Their scaling has only been tested near $T_c$ since  the
upper critical field is beyond the experimental range of observation for low temperature. In this work we apply
the Landau and Ott scaling to $Li_2Pd_3B$ and Nb, which is the most studied of the low $T_c$ materials. We are
motivated by the fact that for the low $T_c$ materials the temperature range covered is
significant larger as compared to the high-Tc materials.\\

The method of Landau and Ott \cite{LO}, inspired in the original Abrikosov's\cite{abrikosov} theory, assumes a
temperature independent Ginzburg-Landau parameter $\kappa$. The basic hypothesis is that the temperature
dependence of the magnetic susceptibility $\chi(H,T)\equiv M(H,T)/H = \chi(h)$ is all contained in the reduced
field $h=H/H_{c2}(T)$, that is, in the upper critical field at temperature T: $H_{c2}(T)$. From this it follows
that,
\begin{equation}
M(H,T)=H_{c2}(T)h\chi(h), \label{eq1}
\end{equation}
and the scaling relation connecting magnetization values at two different temperatures, $T_0$ and $T$, which is,
\begin{equation}
M(H,T_{0})=M(h_{c2}H,T)/h_{c2}, \label{eq2}
\end{equation}
$h_{c2}=H_{c2}(T)/H_{c2}(T_{0})$. This relation implies that for all temperatures the $M(H)$ curves collapse
into a single curve, once a suitable choice of the parameter $h_{c2}(T)$ is chosen for each curve. The collected
set of scaling parameters $h_{c2}(T)$ plotted versus $T$ leads to the determination of $T_c$. Thus the phase
diagram border line, $H_{c2}vsT$, follows from a single known value of $H_{c2}(T_{0})$. Notice that the critical
temperature $T_c$ is not used as a fitting parameter in the method, instead is obtained from extrapolation to
zero field of the $H_{c2}vsT$ scaling curve. A caveat is that all sources of magnetization
other than the superconducting one must be removed before applying the Landau and Ott scaling.\\

In this paper we show that in the reversible region of large $\kappa$ superconductors, the average kinetic
energy density of the condensate is a universal function because its temperature dependence is in the reduced
field $h$. The argument relies on the fact that the kinetic energy density of the condensate divided by the
square of the applied field, $K/H^2$ is a function of the magnetic susceptibility $\chi(h)$. Thus the kinetic
energy density obeys a scaling law that directly follows from the Landau and Ott argument:
\begin{equation}
\frac{K}{H^2}=[1+d.\chi(h)].\chi(h), \label{eq3}
\end{equation}
The susceptibility determines the average kinetic energy density according to the above equation, though this
direct connection is not explored here any further. Instead the average kinetic energy density is obtained
through the equation below. A few years ago some of us\cite{mauro2} have proposed  a new way to plot $MvsH$
isothermal magnetic data, through the $d.M.BvsH$ diagram, $B=(H+d.M)$, $B$ being the magnetic induction, and
$d$, the demagnetization factor of the sample. Remarkably this quantity gives the average kinetic energy of the
condensate for a sufficiently large $\kappa$ superconductor,
\begin{equation}
K=-d.M.B, \label{kdmb}
\end{equation}
as shown through the Virial Theorem\cite{mauro,mauro2}. This new plot unveils remarkable features present in the
isothermal magnetization data. As discussed in Ref.~\onlinecite{mauro} the isothermal plot $d.M.BvsH$ vanishes
in the Meissner region and also above the upper critical field $H_{c2}$, while a minimum (maximum value of the
kinetic energy) is predicted to occur for a specific field $H^*$\cite{mauro}. According to the Ginzburg-Landau
theory this critical field, defined as $H^*$\cite{mauro}, can be directly used to determine $H_{c2}$.\\

The recent discovery of superconductivity in the cubic perovskite $Li_2Pd_3B$ \cite{Togano} has brought renewed
interest in ternary borides containing alkaline and transition metals. Theoretical arguments indicate that
strong electronic correlations \cite{Sardar} are important in this compound, though its low temperature specific
heat behavior, namely its $\gamma$ coefficient, is in agreement with a conventional Fermi liquid\cite{Takeya}.
The $Li_2Pd_3B$ cubic structure is composed by distorted $Pd_6B$ octahedrons resembling the well known
octahedral oxygen structure found in the high-tc superconductors. Magnetic measurements have been performed in
$Li_2Pd_3B$ \cite{Badica} revealing a classic intermetallic compound. The coherence and London penetration
lengths were found to be, $\xi=9.1$ nm and $\lambda=194$ nm, yielding a Ginzburg-Landau parameter $\kappa=21$.
Properties of the phase diagram $HvsT$ were obtained\cite{Badica}, the Kramer curve $H_k(T)$ and the
irreversibility line $H_{irr}(T)$, defined for vanishing critical current, were found to be linear.\\

For Nb, isothermal magnetization curves were obtained here for several temperature values using the same sample
studied in Ref.~\onlinecite{mauro,said2}, whose Ginzburg-Landau parameter is $\kappa=4$. According to
Ref.~\onlinecite{mauro} for $\kappa \ge 3$ the product $d.M.B$ represents the average kinetic energy with
precision better than $1\;\%$. Concerning the reversible (thermodynamic) region, new data points were added to
the curves, leading to a better resolution than those described in Ref.~\onlinecite{mauro,said2}. Points along
these curves were obtained by increasing the field for fixed $\Delta H$ increments at the fixed previously
selected temperatures. In this way a more detailed information about the reversible part of the $HvsT$ diagram
(used in the Landau and Ott analysis) is obtained. Magnetization data were always taken after cooling the sample
in zero field. The measurements were done on a commercial Quantum Design SQUID magnetometer with 3 cm scans. We
also consider here isothermal $MvsH$ hysteresis data taken on irreversible regime.  This data is the same one of
Ref.~\onlinecite{mauro,said2}, here used for a different purpose, namely, the scaling
analysis performed on the $d.M.BvsH$ curves.\\

In this paper we find that the reversible magnetization of $Li_2Pd_3B$ and of Nb obey Landau and Ott scaling.
The original Landau and Ott proposal of scaling was for the reversible magnetization, but later they observed
that scaling also applies below the irreversibility line for increasing fields\cite{LO3}. The product $d.M.B$
describes the average kinetic energy density of large $\kappa$ superconductors, but just for the reversible
region since the pinning interaction is not considered in this connection~\cite{mauro}. Similarly we apply
scaling to the $d.M.B.$ curves of $Li_2Pd_3B$ and Nb that are only partially in the reversible region. In case
of $Li_2Pd_3B$ the scaling theory applied to $d.M.B/H^2$ produces an impressive collapse of all curves over the
entire field range of field measurements. However a similar collapse is not observed for Nb. This failure is
blamed on the existence of a pronounced peak effect that the magnetization curves show near $H_{c2}(T)$, which
is also present in the corresponding kinetic energy curves. Despite of this absence of scaling, curves of
$d.B.M/H^2vsH$ for Nb show the same temperature independent maximum value, located just above $H^*$. This
interesting new feature, only observed in case of Nb, appears to be related to the critical magnetic field where
full penetration in the sample takes place.\\

The demagnetization factor, $d$, used to obtain the quantity $K=d.M.B$, for the studied samples were determined
from the magnetization curves by assuming that $B=H+d.M$ must vanish in the Meissner phase. For both compounds,
Nb and $Li_2Pd_3B$, the removal of the background, namely, of the non-superconducting contribution to the
magnetization, is found to be field dependent but not temperature dependent, as expected from Pauli paramagnetism.\\

 The interest in the study of the reversible or thermodynamic part of the phase diagram stems from the
high-$T_c$ materials \cite{bedernoz,blatter}, which exhibit a large and rich reversible region, oppositely to
the conventional low-$T_c$ materials. High-$T_c$ materials display an upward curvature in the irreversibility
line, attributed, among others, to high-field diamagnetic fluctuations \cite{welp,bula,tesa}. The Landau and Ott
definition of the upper critical field $H_{c2}(T)$ brings a new switch to this controversial issue as it
displays no upward curvature. We determine meaningful $H_{c2}$ values for $Li_{2}Pd_{3}B$ and Nb using the
Landau and Ott scaling method on the reversible magnetization.
data.\\

\begin{figure}
\includegraphics[width=1.0\linewidth, height=0.7\linewidth]{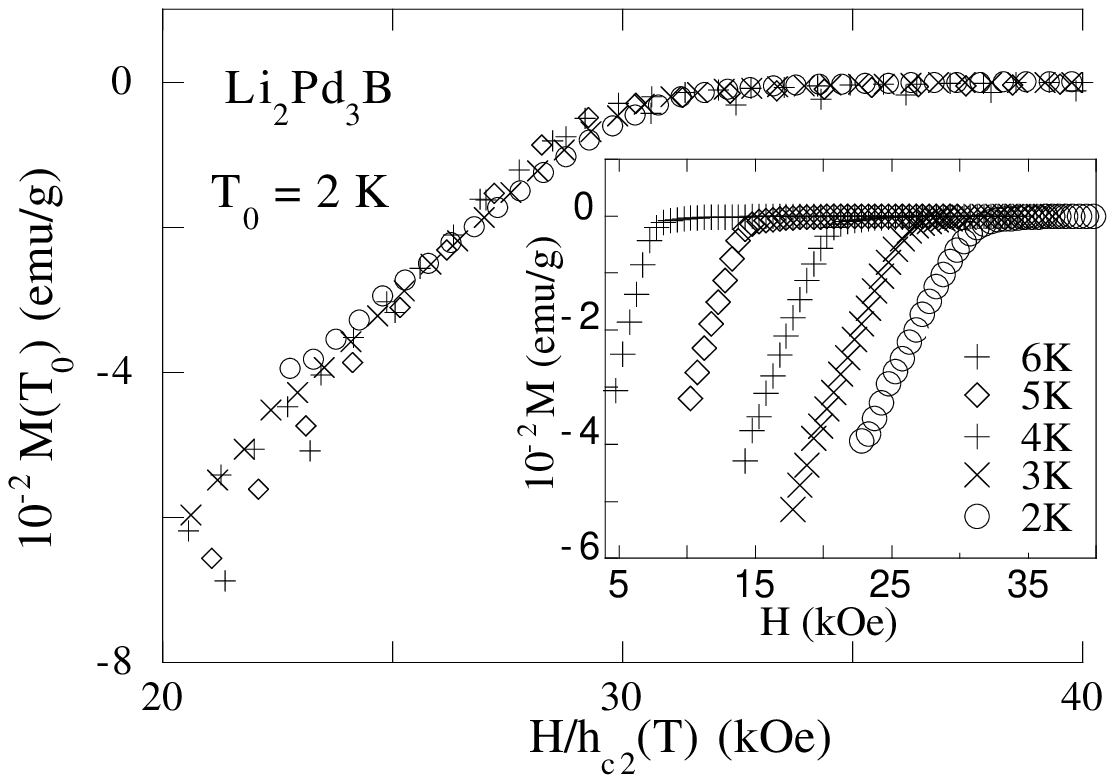}
\caption{The collapse of several $Li_2Pd_3B$ isothermal reversible magnetization curves to the $T_0=2$K curve,
according to the Landau-Ott scaling. The inset shows the set of magnetization curves and their corresponding
temperatures.} \label{fig1}
\end{figure}

Fig.~\ref{fig1} shows the scaling analysis performed on the $Li_{2}Pd_{3}B$, according to  Eq.~\ref{eq2}. The
collapse of the different temperature was achieved by searching suitable values of $h_{c2}$, following the
choice of $T_{0}=2K$. We observe that the collapse only occurs for the reversible region, whose universal
magnetization curve is shown in  Fig.~\ref{fig1}.\\

\begin{figure}
\includegraphics[width=1.0\linewidth, height=0.7\linewidth]{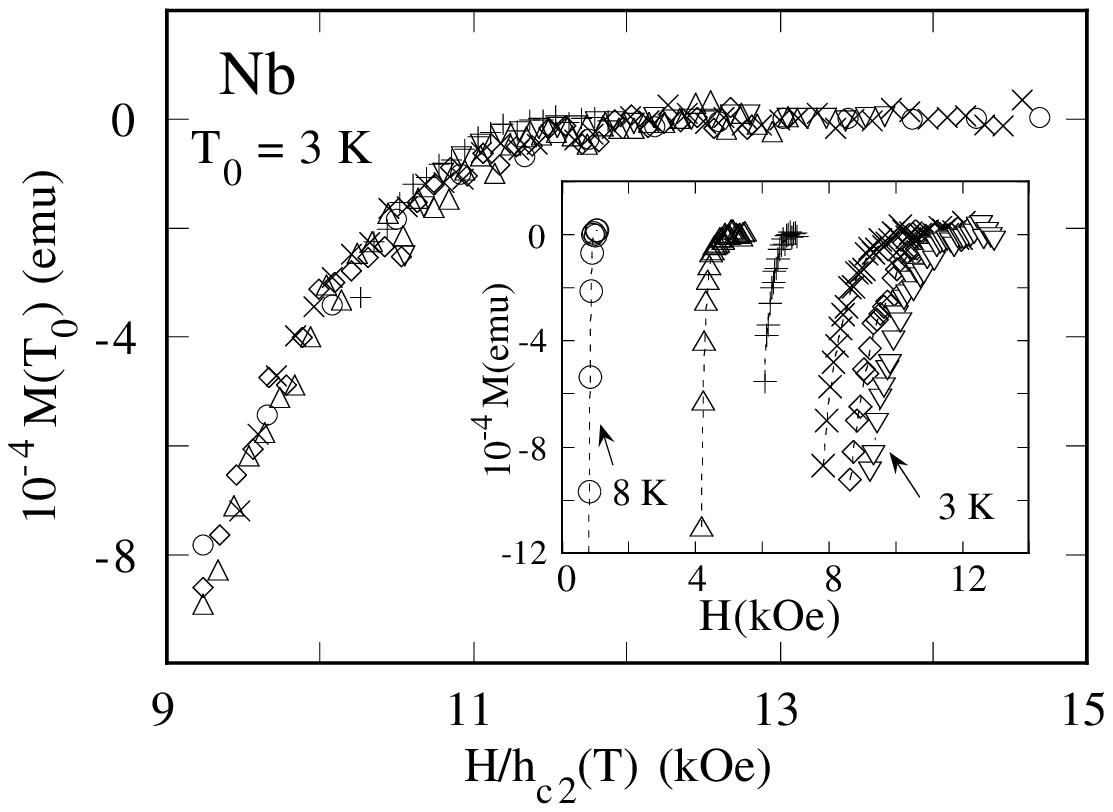}
\caption{The collapse of several Nb isothermal reversible magnetization curves to the $T_0=3$K curve, according
to the Landau-Ott scaling. The inset shows the set of magnetization curves and their corresponding
temperatures.} \label{fig2}
\end{figure}

Fig.~\ref{fig2} shows the same analysis of Fig.~\ref{fig1} performed on the Nb reversible magnetization data
with the background magnetization already subtracted. The selected temperature for scaling is $T_{0}=3K$.
Concerning the scaling approach shown in Figs.~\ref{fig1} and ~\ref{fig2} notice that the for these low $T_c$
materials the isothermal $MvsH$ curves are more far apart from each other than the similar curves for high-$T_c$
materials \cite{LO,LO2,LO3}.\\

\begin{figure}
\includegraphics[width=1.0\linewidth, height=0.7\linewidth]{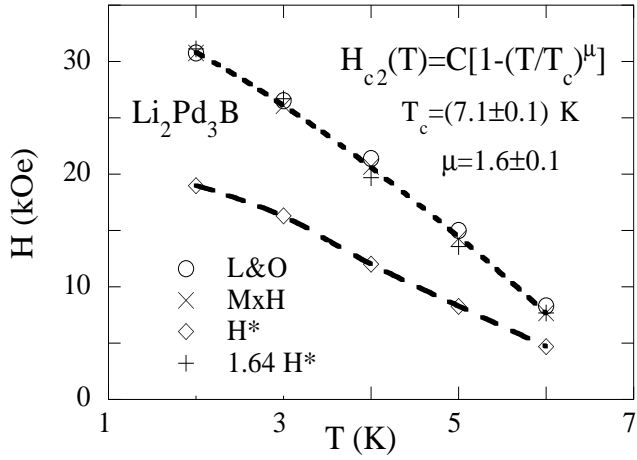}
\caption{Three $H_{c2}(T)$ curves for $Li_2Pd_3B$ are shown here, obtained directly from the $MvsT$ reversible
curves, from the maximum in the kinetic energy curves and from the Landau-Ott scaling method. The curve $H^*(T)$
of maximum kinetic energy is also shown here.} \label{fig3}
\end{figure}

\begin{figure}
\includegraphics[width=1.0\linewidth, height=0.7\linewidth]{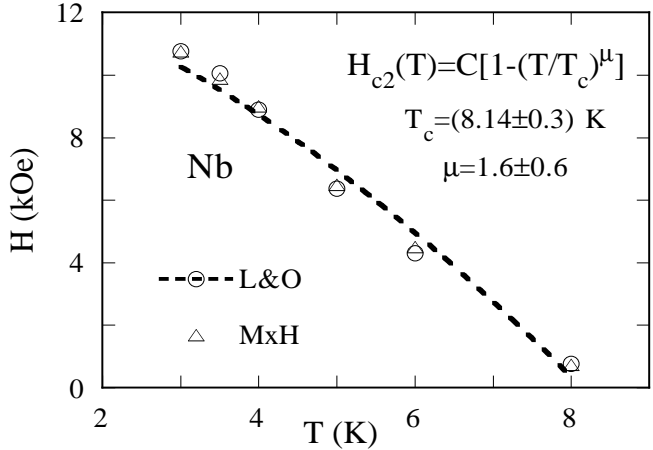}
\caption{Two $H_{c2}(T)$ curves for Nb are shown here, obtained directly from the $MvsT$ reversible curves and
from the Landau-Ott scaling method. From this last method one obtains the plotting exponent $\mu$ given here.}
\label{fig4}
\end{figure}

The phase diagrams of Figs.~\ref{fig3} and ~\ref{fig4} show $H_{c2}(T)$ values  for temperatures much below
$T_c$, away from the linear regime of the phase diagram. The fitting shown in Figs.~\ref{fig3} and ~\ref{fig4}
was obtained with $H_{c2}(T)=C[1-(T/T_{c})^{\mu]}$, where $\mu=1.6$ for both studied samples and  produced
reasonable values for $T_c$. Fig.~\ref{fig3} shows the phase diagram for $Li_{2}Pd_{3}B$. For Landau and Ott
scaling $H_{c2}(T)$ is obtained from the $h_{c2}$ values of the collapsed curve of Fig.~\ref{fig1}, multiplied
by $H_{c2}(T_{0})$ obtained from the Abrikosov method. Fig.~\ref{fig3} also displays the values of $H_{c2}(T)$
obtained from the linear extrapolation of the reversible region till $M=0$. Again, a remarkable agreement
between both values of $H_{c2}(T)$ is found. The maximum of the kinetic energy, $H^*$, for $Li_{2}Pd_{3}B$ was
extracted from the $d.M.BvsH$ curves and plotted in Fig.~\ref{fig3} and the constant ratio
$H^{*}(T)/H_{c2}(T)\approx 0.61$ was found valid, thus nearly temperature independent in the studied range. The
same temperature independent ratio was previously found to be $0.5$ for our Nb sample \cite{mauro}. To exemplify
this technique, we also plot in Fig.~\ref{fig3} values of $H_{c2}(T)$ obtained from the ratio
$H_{c2}(T)/H^{*}(T)\equiv C$ where the constant $C=0.61$ is obtained from the ratio $H_{c2}(T=6K)/H^{*}(T=6K)$.
As shown in Fig.~\ref{fig3} this is in good agreement with the Landau and Ott scaling and direct $MvsH$ methods.
The later method might be of particular interest when studying systems with higher values of $H_{c2}(0)$, since
values of $H_{c2}(T)$ which may not be achieved experimentally can be
estimated from the much lower values of $H^*$.\\

Fig.~\ref{fig4} shows the phase diagram obtained for Nb. The obtained values of $h$ added to the value of
$H_{c2}(T_{0})$ determined from the Abrikosov method. Fig.~\ref{fig4} also displays values of $H_{c2}(T)$
obtained for the same curves presented in the inset of Fig.~\ref{fig2}, but using the Abrikosov method. It is
remarkable that the later values of $H_{c2}(T)$ for Nb
virtually coincides with the values obtained from  the scaling analysis of Fig.~\ref{fig2}.\\

\begin{figure}
\includegraphics[width=1.0\linewidth, height=0.7\linewidth]{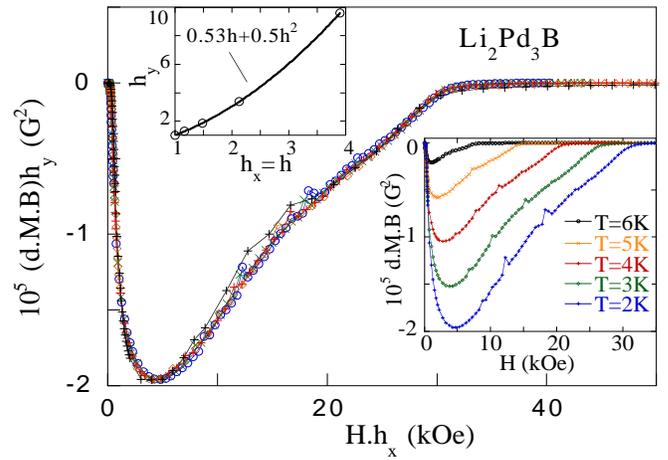}
\caption{The collapse of several $Li_2Pd_3B$ isothermal kinetic energy curves after scaling, following
Eqs.~\ref{eq3}and ~\ref{kdmb}. The lower inset shows the set of isothermal curves used in the main figure. The
upper inset shows the scaling factors used to obtain the collapsed curve.} \label{fig5}
\end{figure}
Fig.~\ref{fig5} shows the scaling analysis for $Li_2Pd_3B$. The inset of Fig.~\ref{fig5} shows 2K to 6 K $MvsH$
curves extracted from Ref.~\onlinecite{Togano}. Their collapse into a single curve is impressive and extends
over the entire range of field measurements, which includes the irreversible region. The universal curve shown
in Fig.~\ref{fig5} was obtained after multiplying the y-axis and the x-axis of the $d.M.B/H^2$ curves, shown in
the lower inset of this Figure, by $h_y(T)$ and $h_x(T)$ where the values and the relation between $h_y(T)$ and
$h_x(T)$ is presented in the upper inset of Fig.~\ref{fig5}. It is worth mentioning that the relation between
$h_y(T)$ and $h_x(T)$ does not spoil the $h$ scaling as $h_x(T)$ virtually coincides with $h(T)$ obtained after
applying the Landau and Ott scaling to the reversible magnetization curves. We stress that the relation between
$h_y(T)$ and $h_x(T)$ is a fitting result found to be close to the Eq.~\ref{eq3} prediction.\\

\begin{figure}
\includegraphics[width=1.0\linewidth, height=0.7\linewidth]{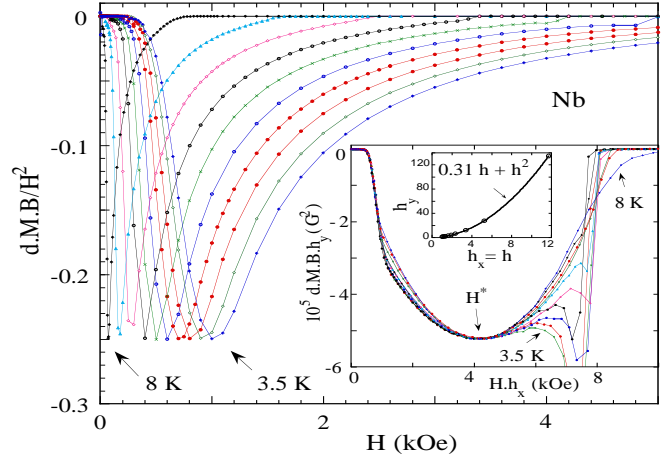}
\caption{Isothermal kinetic energy curves, $d.M.B/H^2vsH$, for Nb. The inset shows the collapse of the
$d.M.BvsH$ curves, shows in the main figure, after scaling following Eqs.~\ref{eq3}and ~\ref{kdmb}. The inner
inset shows the scaling factors used to obtain the collapsed curve.} \label{fig6}
\end{figure}
Fig.~\ref{fig6} shows curves of the quantity $d.M.B/H^2$ for Nb, extracted from data obtained in
Ref.~\onlinecite{mauro}. The lower inset of Fig.~\ref{fig6} shows the results of the kinetic energy scaling
applied to the Nb curves. As mentioned in Fig.~\ref{fig5} above, the collapse shown in the inset of
Fig.~\ref{fig6} covers the full region of field measurement, including the irreversible region. The inner inset
display values of the scaling factors $h_y(T)$ and $h_x(T)$ used to obtain the lower inset figure. Notice that
the relation between $h_y(T)$ and $h_x(T)$ is a best fitting choice similar to the Eq.~\ref{eq3} prediction. The
arrow in the inset shows the position of $H^*$, and $h_x=H^*(T=3.5K)/H^*(T)$ and $h_y=K^*(3.5K)/K^*(T)$, where
$K^*=K(H^*)$. As in Fig.~\ref{fig5}, the relation between the scaling factors are in agreement with the scaling
hypothesis. The values of $h_x(T)$ are in agreement with the scaling factor values $h(T)$ obtained after
applying the Landau and Ott scaling to the reversible magnetization curves shown below. One sees in the lower
inset of Fig.~\ref{fig6} that the collapsed kinetic energy curves mainly differ for fields above the minimum
$d.M.B.$ point, namely for $H>H^*$. As previously mentioned this lack of scaling is related to the peak effect.
An interesting result arises despite of this lack of scaling. All $d.M.B/H^2$ curves exhibit the same minimum
value which appears to be an intrinsic temperature independent effect. We mention that the field position of
each minimum occurs very close to the field for which the sample is fully penetrated by the external field. This
view is supported by the comparison between the $d.M.BvsH$ plot and the corresponding $MvsH$ curve. A similar
plot for $Li_2Pd_3B$ does not show a temperature independent maximum value, probably due to surface barriers or,
and, edge effects in this sample not considered in the present scaling approach developed for the bulk. One may
observed that the curves of Fig.~\ref{fig6} suggests a new collapse of all $d.M.B/H^2$ curves by fine tuning
x-axis factors so to adjust this secondary maximum to the same position. However this proposal does not work due
to the pronounced peak effect, already  present in our $MvsH$ curves, and enhanced in the $d.M.B./H^2$ curves.\\

In conclusion the reversible magnetization of $Li_2Pd_3B$ and Nb are shown here to satisfy the Landau and Ott
scaling as shown in Figs.~\ref{fig1} and ~\ref{fig2}. We also considered isothermal magnetization measurements
done in zero field cooled $Li_2Pd_3B$ and Nb samples and used previous data obtained in increasing field data.
For large $\kappa$ superconductors we claim universality of the average kinetic energy of the condensate based
on the virial theorem, but Eqs.~\ref{eq3} and ~\ref{kdmb} were derived for the reversible regime. Zero field
cooled isothermal magnetization data, casted as the product $d.M.B./H^2$ shows impressive scaling for
$Li_2Pd_3B$ but not for Nb. Interestingly the Nb isothermal $d.M.B/H^2$ curves exhibit a constant and
temperature independent minimum value. The upper critical curve of the superconductor $Li_{2}Pd_{3}B$ is
obtained based on two new methods, namely, the Landau and Ott scaling and the maximum kinetic energy field. For
comparison we have also studied Nb in order to display similar features in the low temperature compounds.\\

We thank CNPq, FAPERJ and one of us (mmd) also thanks the Instituto do Mil\^enio de Nanoci\^encias for financial support.\\
$\ast$ Corresponding author. E-mail: mmd@if.ufrj.br

\end{document}